\documentclass[twocolumn,]{aastex631}
\usepackage{amsmath}

\makeatletter
\newcommand\footnoteref[1]{\protected@xdef\@thefnmark{\ref{#1}}\@footnotemark}  
\makeatother

\newcommand{\sersic}{S\'ersic}
\newcommand{\lenstronomy}{\texttt{lenstronomy}}
\newcommand{\galight}{\texttt{galight}}
\newcommand{\reff}{{$R_{\mathrm{eff}}$}}
\newcommand{\smass}{{$M_*$}}
\newcommand{\hst}{{\it HST}}
\newcommand{\jwst}{{\it JWST}}
\newcommand{\angstrom}{\text{\normalfont\AA}}

\begin{document}
\title{Opening the era of quasar host studies at high redshift with JWST}

\author[0000-0002-0786-7307]{Xuheng Ding}
\email{xuheng.ding@ipmu.jp}
\affiliation{Kavli Institute for the Physics and Mathematics of the Universe, The University of Tokyo, Kashiwa, Japan 277-8583 (Kavli IPMU, WPI)}

\author[0000-0002-0000-6977]{John D. Silverman}
\affiliation{Kavli Institute for the Physics and Mathematics of the Universe, The University of Tokyo, Kashiwa, Japan 277-8583 (Kavli IPMU, WPI)}
\affiliation{Department of Astronomy, School of Science, The University of Tokyo, 7-3-1 Hongo, Bunkyo, Tokyo 113-0033, Japan}

\author[0000-0003-2984-6803]{Masafusa Onoue}
\affiliation{Kavli Institute for Astronomy and Astrophysics, Peking University, Beijing 100871, China}
\affiliation{Kavli Institute for the Physics and Mathematics of the Universe, The University of Tokyo, Kashiwa, Japan 277-8583 (Kavli IPMU, WPI)}

\begin{abstract}
We measure the host galaxy properties of five quasars with $z\sim 1.6 - 3.5$ selected from SDSS and AEGIS, which fall within the \jwst/\hst\ CEERS survey area. A PSF library is constructed based on stars in the full field-of-view of the data and used with the 2-dimensional image modeling tool \galight\ to decompose the quasar and its host with multi-band filters available for \hst\ ACS+WFC3 and \jwst\ NIRCAM (12 filters covering \hst\ F606W to \jwst\ F444W). As demonstrated, \jwst\ provides the first capability to detect quasar hosts at $z>3$ and enables spatially-resolved studies of the underlying stellar populations at $z\sim2$ within morphological structures (spiral arms, bar) not possible with \hst. Overall, we find quasar hosts to be disk-like, lack merger signatures, and have sizes generally more compact than typical star-forming galaxies at their respective stellar mass, thus in agreement with results at lower redshifts. The fortuitous face-on orientation of SDSSJ1420+5300A at $z = 1.646$ enables us to find higher star formation and younger ages in the central $2-4$ kpc region relative to the outskirts, which may help explain the relatively compact nature of quasar hosts and pose a challenge to AGN feedback models.

\end{abstract}
\keywords{Galaxy evolution(594) --- AGN host galaxies(2017) --- Quasars (1319)  --- Active galaxies(17)}

\section{Introduction} \label{sec:intro}

A key open question in astrophysics is how the tight correlations between the mass of supermassive black holes (SMBH) and their host galaxy properties (e.g.,~velocity dispersion, bulge stellar mass) emerge~\citep[e.g.,~][]{Mag++98, F+M00, M+H03, H+R04, Gul++09, Geb++01b, Gra++2011,Beifi2012}. Galaxy mergers are too rare, and most quasar hosts do not show signs of interactions or disturbances to be a primary driver \citep[e.g.,][]{Cisternas2011,Kocevski2012,Mechtley2016}. Thus, secular (i.e., internal) processes are the only other conceivable options that can concurrently feed the black hole and build the central stellar mass concentration. Likely relevant, quasars hosts have been recently shown to have sizes (effective radius) more compact than the star-forming population but not yet as small as the quiescent population \citep{Silverman2019,Li2021}. We do not yet know whether those central stars are formed in situ or brought in from further out, possibly due to minor mergers.

To make further progress, a decomposition of the host galaxy from its central quasar is required to measure the host properties at earlier cosmic times ($z\gtrsim2$) when the central regions of galaxies are forming stars and particularly those that can be spatially resolved. At cosmological distances, galaxies have smaller apparent sizes and become fainter due to surface brightness dimming. The unresolved quasars, seen as bright point sources, can outshine the host galaxy, which makes the decomposition of the host challenging, even with {\it Hubble Space Telescope} (\hst) WFC3-IR, which has excelled at $z < 2$. Owing to this reason, most of the studies in the literature so far have been focused on the moderate-luminosity AGNs whose brightness is comparable to their host ~\citep{Park15, Tre++07, Bennert11, Woo++08, Jah++09, SS13, Mechtley2016,Ding2020, Ding2021, 2021ApJ...906..103L}. Hence, the stellar content of the luminous AGN hosts at $z>2$, which harbors the most massive SMBHs, has not yet been studied. 

The {\it James Webb Space Telescope} (\jwst) opens up a new opportunity to extend the study of quasar hosts at high-$z$. The infrared coverage at $\lambda > 2 \mu$m provides the ability to probe the galaxy 4000~\angstrom\ break beyond $z\sim 3.5$.  The unprecedented sensitivity and resolution by \jwst\ is key to separating the unresolved quasar component from the extended emission of its host galaxy, particularly due to the larger aperture compared to \hst\ and the stability of the telescope, which effectively improves the characterization of the PSF. 

In this paper, we use deep \textit{Near Infrared Camera} \citep[NIRCam,][]{Rieke05} imaging data from the \jwst\ Cosmic Evolution Early Release Science\footnote{\url{https://ceers.github.io}}~\citep[CEERS,][]{CEERS} program to decompose the optical and infrared emission from five quasars with redshift between 1.6 and 3.5. We limit the AGN sample to those with spectroscopic redshifts to provide accurate stellar masses and sizes of their host galaxies. We carry out a quasar decomposition to reveal the host galaxy and then perform the spectral energy distribution (SED) fitting to infer the stellar mass, star formation rate (SFR), and stellar age. These results will aid in our ability to establish relations between AGN and host galaxy formation with \jwst\ \citep{Kocevski2022}.

The paper is organized as follows. In Section~\ref{sec:design}, we describe the \jwst/\hst\ data and our sample selection. We introduce our decomposition method in Section~\ref{sec:decomp}, including our PSF library construction, decomposition tool, and SED fitting. In Section~\ref{sec:result}, we present our measurements and compare to the size - mass relations of non-active galaxies from the literature. The concluding remarks are presented in Section~\ref{sec:conclusion}. Magnitudes are given in AB System. A Chabrier initial mass function is employed to measure the stellar mass of the host galaxies. We use a standard concordance cosmology with $H_0= 70$ km s$^{-1}$ Mpc$^{-1}$, $\Omega{_m} = 0.30$, and $\Omega{_\Lambda} = 0.70$.

\section{Experimental design} \label{sec:design}

\subsection{CEERS survey imaging and data reduction}
The CEERS \jwst\ early release science program (ERS Program \#1345) is designed to cover 100 arcmin$^2$ of the AEGIS fields with imaging and spectroscopic data~\citep{CEERS}. This program will observe ten pointings with \jwst/NIRCam, six with NIRSpec in parallel, and four with MIRI in parallel. In late June 2022, four pointings  (CEERS1, CEERS2, CEERS3, and CEERS6) were completed using 7 NIRCam filters: F115W, F150W, F200W, F277W, F356W, F410M, and F444W. The integration times are 2,635 seconds for each filter, with the exception of F115W, which was taken with an exposure time twice as long. For the CEERS2 pointing, one additional visit was performed with a slightly different PA and positional offset using filters F200W and F444W. We refer the reader to~\citet{Finkelstein2022} and the CEERS overview paper (Finkelstein et al. in prep) for more details on the \jwst\ observations and survey design including the positions of the three dither pattern. This information can also be found via the\dataset[DOI]{https://doi.org/10.17909/3pf0-8b20} information.

\begin{figure*}
\centering
\includegraphics[trim={3cm 0cm 3cm 0cm},clip,width=1.8\columnwidth]{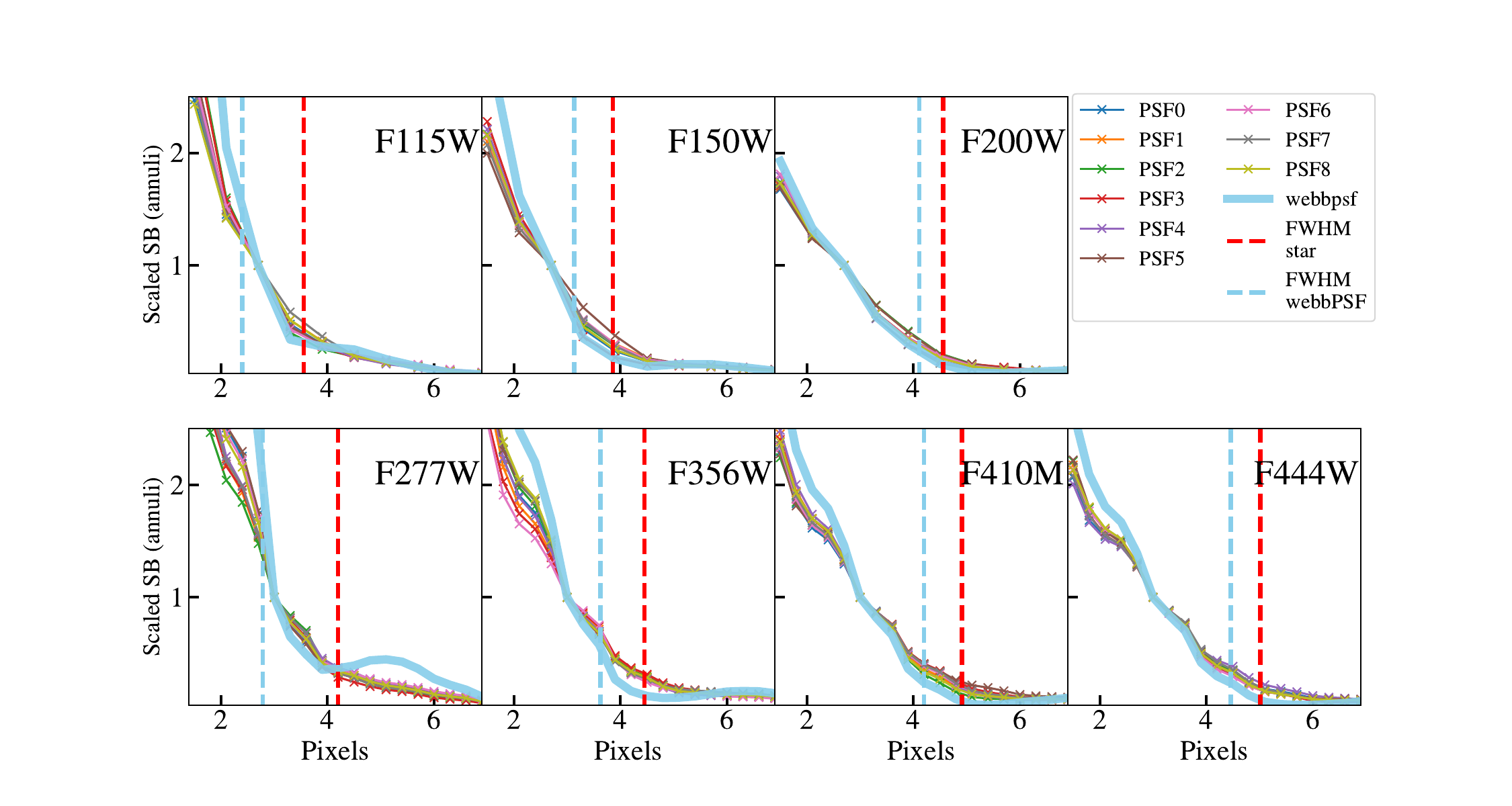}
\caption{Surface brightness (SB) profiles (annuli) of nine select PSFs are shown as a function of radius (based on stars detected by \jwst) per filter and compared with the simulated PSF model from \texttt{webbpsf}. The vertical lines indicate the FWHM values of the PSF-stars (averaged) and PSF model by \texttt{webbpsf}. We only show the inner regions ($<7$ pixels) of these profiles which are key for quasar subtraction. The PSFs in our library demonstrate very stable performance, while the simulated PSF models by \jwst\ are too narrow (i.e., peaking at the center) for all the filters.
\label{fig:PSFprofile}}
\end{figure*}

 We performed our own data reduction to achieve science-quality images. The archived data products (stage 2) were acquired from the STScI MAST Portal\footnote{\url{https://archive.stsci.edu/}}. The stage 2 data were processed by the \jwst\ pipeline version 1.5.3 with the mapping file \textsf{jwst\_0942.pmap} except for F410M and F444W. Depending on the fields, we use \textsf{jwst\_0877.pmap} and \textsf{jwst\_0878.pmap} for F410M. For F444W, we use \textsf{jwst\_0877.pmap}, \textsf{jwst\_0878.pmap} and \textsf{jwst\_0881.pmap}. Due to the known issue with the background subtraction for the v.1.5.3 pipeline, we further post-process the stage 2 images as follows. We first subtracted the background light using \texttt{photutils} based on \textsf{Background2D} function. 
 The so-called $1/f$ noise pattern due to detector read noise was removed with the code provided by the CEERS team\footnote{\url{https://ceers.github.io/releases.html\#sdr1}}. 

We then use the \jwst\ stage 3 pipeline to produce the individual images. In the  \textsf{Resample} step, we increase the pixel resolution by a factor of two with the drizzle algorithm. The pixel scale for the long-wavelength (LW) filters is $\sim$ 0\farcs{}0315 and short-wavelength (SW) filters $\sim$ 0\farcs{}0156. Saturated pixels are labeled as empty in this process. For the purposes of building a PSF library (see Section~\ref{sec:psfLibrary}), we require all four CEERS fields to have the same rotation angle. Thus, we do not apply any rotations when drizzling the images. For WCS calibration, we use the GAIA DR3 source catalog\footnote{\url{https://gea.esac.esa.int/archive/}}~\citep{GAIA16} to align the images. 

The conversion factor from the instrumental signal to physical units Mega-Janskys per steradian is based on the calibration files \textsf{jwst\_nircam\_photom\_0101.fits} for module A, and \textsf{jwst\_nircam\_photom\_0104.fits} for module B. This step requires a re-calibration to correct the original flux based on pre-flight measurements. A significant difference ($10\%-20\%$ flux level) between the pre-flight and post-flight photometric reference files is noted and mentioned in~\citet{Adams22}. We have updated our calibration for the F410M and F444W flux densities based on this information. Until today, the calibration file of \jwst\ still requires updating, and there is still up
to 20\% uncertainty in conversion between the ADU and flux in these filters. As a note, this reduced data was also effective in identifying a low-luminosity AGN candidate at $z=5$~\citep{2022arXiv220907325O}.

For the \hst\ imaging, we adopt the datasets provided by CEERS collaboration\footnote{\url{https://ceers.github.io/hdr1.html\#hst-egs}}. These \hst\ images are observed from eight different \hst\ programs (10134, 12063, 12099, 12167, 12177, 12547, 13063 and 13792), with a total of 1,767 exposures. The data reduction follows the procedures described in~\citet{Koekemoer2011}. We adopt the filters including ACS/WFC F606W and F814W, and WFC3/IR F125W, F140W, and F160W. The final pixel scale has been drizzled to 0\farcs{}03 for all bands.

\subsection{Sample Selection}
We searched for quasars in the SDSS DR16 quasar catalog~\citep{2020ApJS..250....8L} and the AEGIS-X AGN catalog~\citep{2009ApJS..180..102L, Nandra2015} that fall within the CEERS field. We also limited our targets to those with spectroscopic redshifts greater than 1.6 to focus on high redshift quasars, which have been challenging for \hst\ to detect their host. Three SDSS quasars were identified with redshifts at 1.646, 2.588, and 3.442, respectively. SDSS1419+5254 ($z=3.442$) falls in CEERS2 and thus has only two \jwst\ filters (i.e., F200W and F444W) available at the time. From AEGIS-X, we found four quasars with spectroscopic redshifts above $z>1.6$ with two of them matched to SDSS quasars\footnote{We find SDSS1419+5254 and AEGIS 585 are the same system;  SDSS1420+5300A and AEGIS 742 are the same system.}. As a result, two additional targets are selected with redshift at 2.317 and 3.465, respectively. While MIRI images were taken in parallel, no additional quasars were found. We note that only one quasar (AEGIS 482) is in common with a recent \jwst\ study of X-ray-selected AGN in CEERS \citep{Kocevski2022}. The properties of our five quasars are given in Table~\ref{table:sample}.

\section{Quasar-host image decomposition} \label{sec:decomp}
We use our previous 2D image modeling strategy, as fully presented in~\citet{Ding2020} and originally based on HST/WFC3 imaging of quasars at $z\sim1.5$, to perform the AGN decomposition. We first build a PSF library by searching for all point sources (i.e., stars) in the \jwst\ field of views (FOVs). We then perform the fitting of the main quasar target by using the 2D image modeling tool \galight~\citep{Ding2020, Ding2021a} and consider each PSF-star in the library. The final PSF will be chosen from the top-ranked fits as described below.

\subsection{PSF Library}\label{sec:psfLibrary}
The key to AGN-host decomposition is the quality of the PSF, especially when revealing the host galaxy for luminous quasars whose central point source usually dominates the emission. The PSF shape varies with color, brightness, position across the detector, and over time due to aberration and telescope breathing.

We use the previous strategy 
and collect all isolated, unsaturated PSF stars with sufficient signal-to-noise levels to build our PSF library. We use the PSF-pickers (\texttt{find$\_$PSF()} function) provided by \galight\ (see section~\ref{subsec:decomp} for details) to help us collect these objects having a point-like feature as our initial PSF candidates through all the \jwst/\hst\ filters in four CEERS visits. By sorting the FWHM values from low to high and manually viewing the local image stamps of each PSF candidate, we select the PSF candidates that are isolated and sharp. To build the library, we select at least 12 PSF stars for each filter.  Note that these PSFs are normalized to unity by \galight\ before fitting. We provide the positions of each object used to construct the PSFs for each filter through an online website\footnote{\url{https://github.com/dartoon/my_code/blob/master/share_data/PSF_library_info.txt}}.

It is also known that simulated PSF for \hst\ based on \texttt{TinyTim}~\citep{2011SPIE.8127E..0JK} is usually inaccurate (narrow) and not ideal for our goal of AGN decomposition~\citep[e.g.,][]{Mechtley2012}. In this work, we also test the effectiveness of \texttt{webbpsf}~\citep{2014SPIE.9143E..3XP} to generate the PSF model for \jwst\ but find that the simulated PSF is also too narrow compared with the empirical PSFs our the library \citep[see][for similar conclusions]{Ono2022}. We show the surface brightness profiles of PSF stars and the PSF by \texttt{webbpsf} as a function of radius (in pixel unit) in Figure~\ref{fig:PSFprofile}.

\begin{figure*}
\centering
\includegraphics[trim={7cm 1.5cm 7cm 0.5cm},clip,width=1.7\columnwidth]{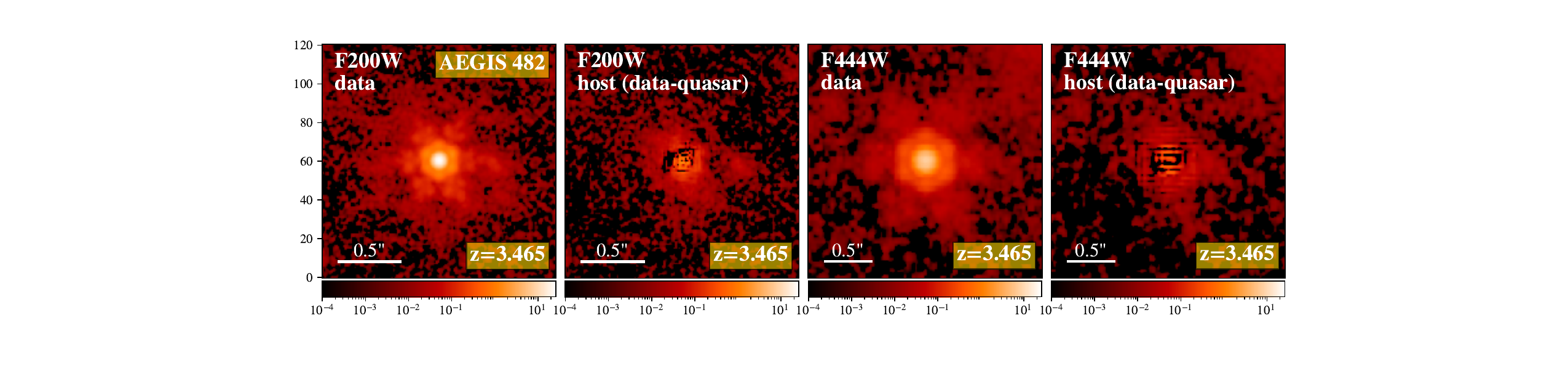}
\includegraphics[trim={7cm 1.5cm 7cm 0.5cm},clip,width=1.7\columnwidth]{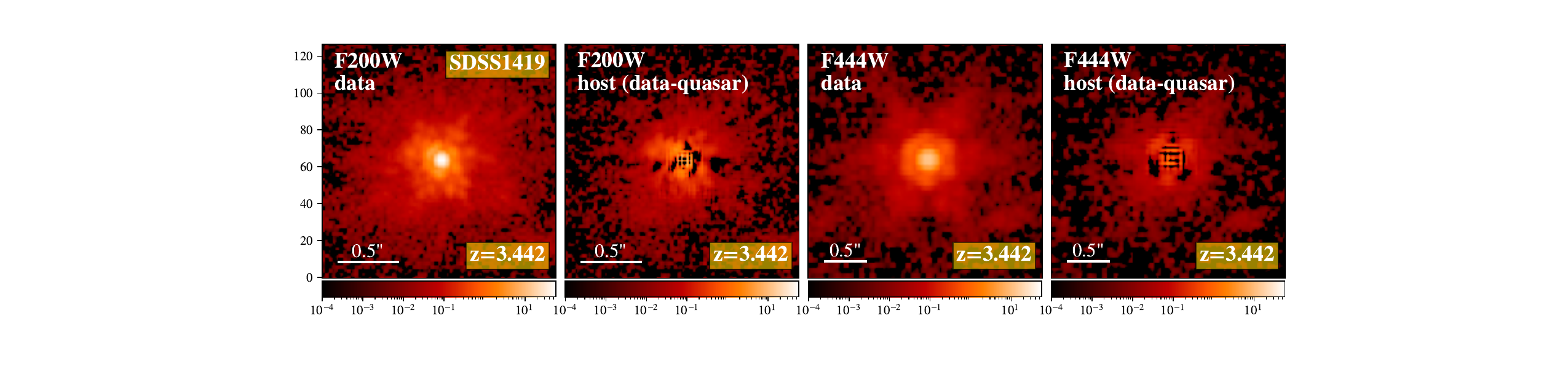}
\includegraphics[trim={7cm 1.5cm 7cm 0.5cm},clip,width=1.7\columnwidth]{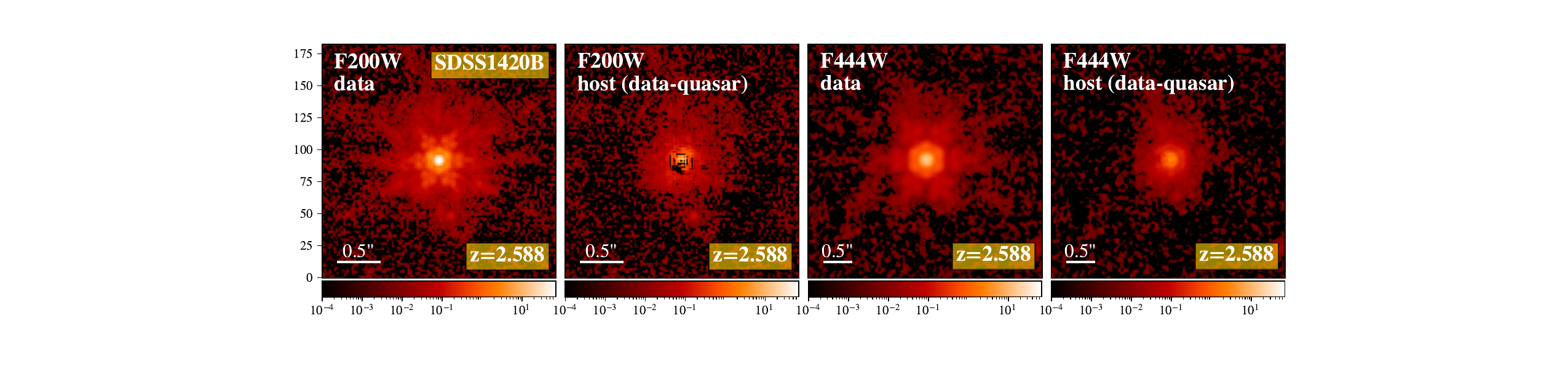}
\includegraphics[trim={7cm 1.5cm 7cm 0.5cm},clip,width=1.7\columnwidth]{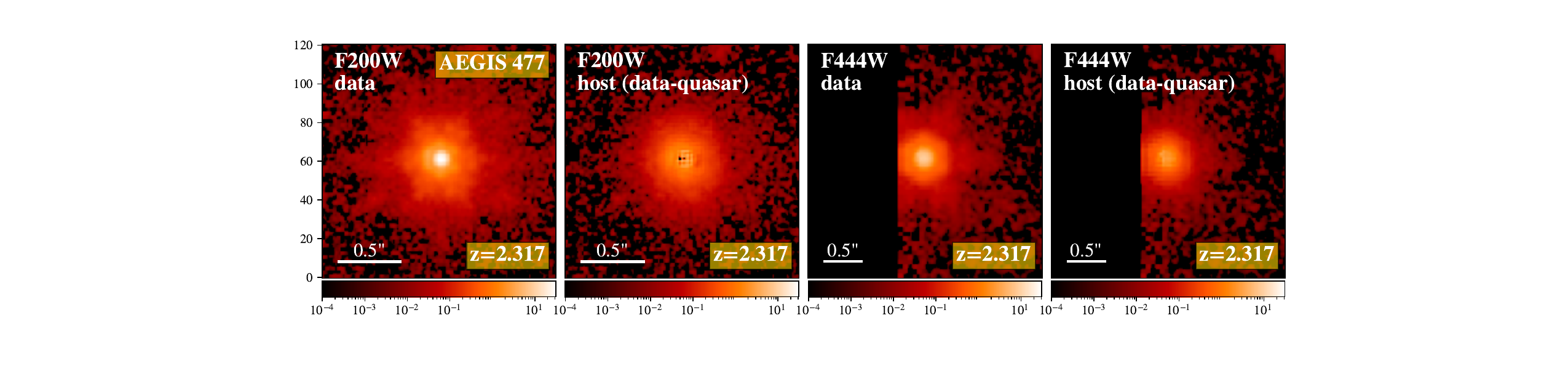}
\caption{Science-grade images (1st and 3rd columns) and host galaxy images (2nd and 4th columns) for the four quasars at $z>2$ with F200W and F444W filters. In each case, the host galaxy image (i.e., data minus quasar) is produced by subtracting the best-fit PSF. The redshift of each quasar is indicated in each panel. The pixel units are in Mega-Janskys per steradian (MJy/sr).
For the LW (e.g., F444W) filter, AEGIS 477 happens to be located at the edge of FOV in module B; only the right-side part of the quasar is observed.}
\label{fig:datahost}
\end{figure*}

\begin{figure*}
\centering
\includegraphics[trim={9cm 1cm 3cm 0cm},clip,width=2\columnwidth]{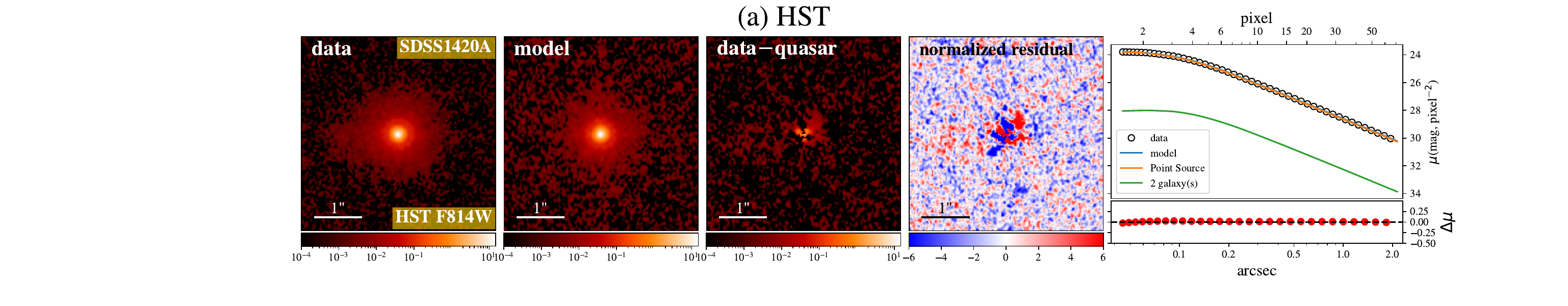}  \\
\vspace*{-0.2cm} 
\includegraphics[trim={9cm 1cm 3cm 0.5cm},clip,width=2\columnwidth]{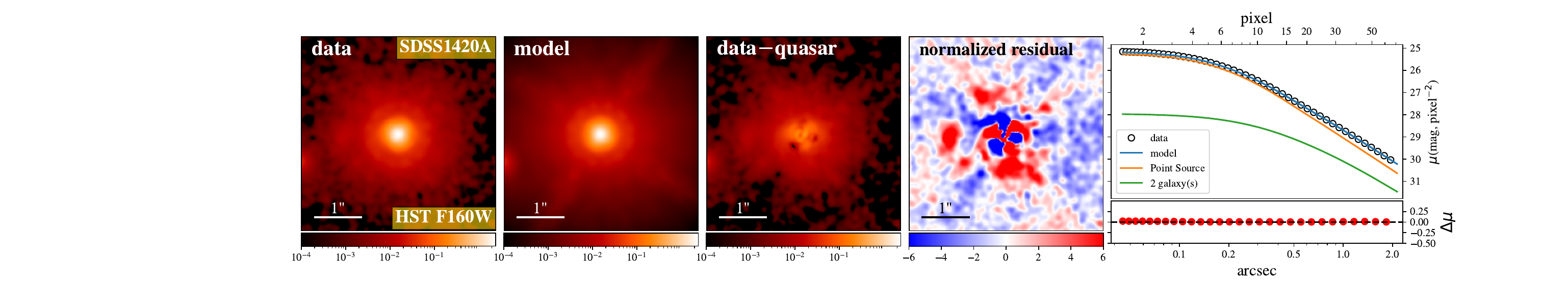}  \\
\vspace*{0.6cm} 
\includegraphics[trim={9cm 1cm 3cm 0cm},clip,width=2\columnwidth]{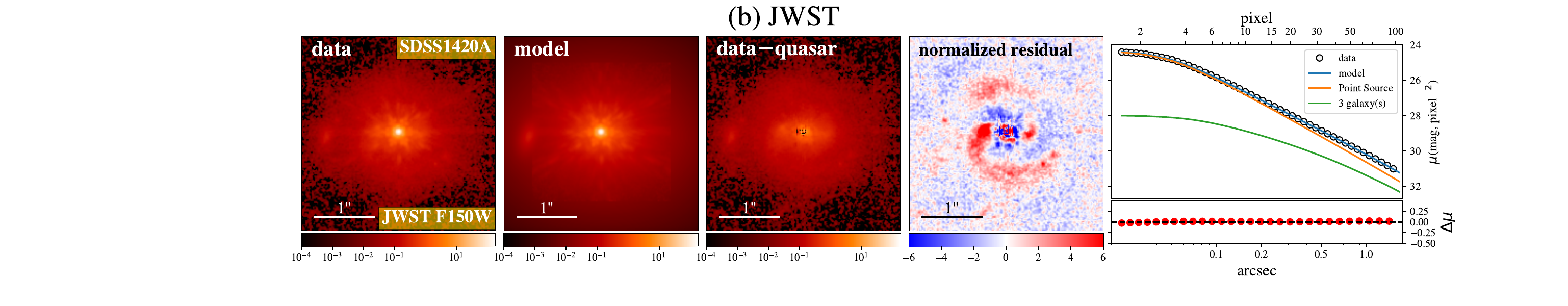}  \\  
\vspace*{-0.2cm} 
\includegraphics[trim={9cm 1cm 3cm 0.5cm},clip,width=2\columnwidth]{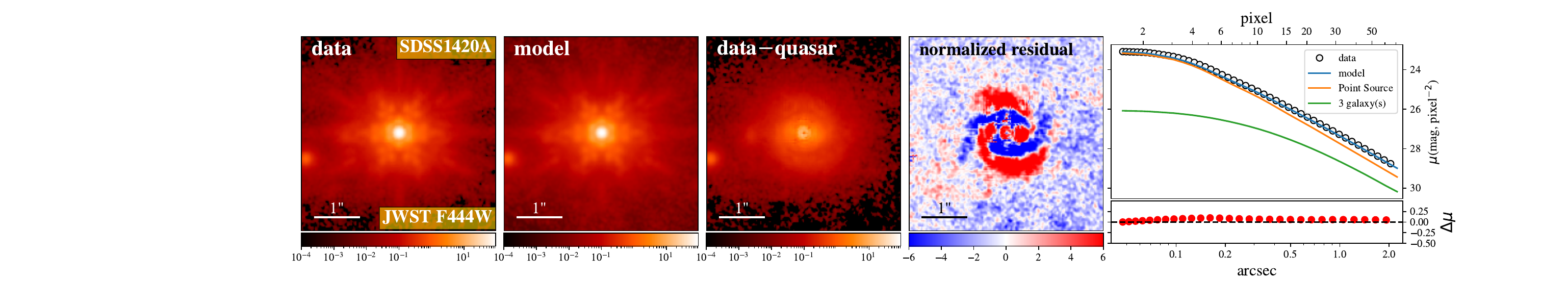}  \\
\caption{2D decomposition of SDSS1420A based on imaging data from \hst\ in panel~(a) and \jwst\ in panel~(b); two filters are selected for each telescope. The panels are as follows from left to right: original data, model (quasar + galaxies) convolved with the PSF, data $-$ quasar (host galaxy only), normalized residual image, and 1D surface brightness profile (top), and the corresponding residual (bottom).
\label{fig:fitting}}
\end{figure*}

\subsection{Image Decomposition}\label{subsec:decomp}

Two-dimensional modeling of images is performed based on our self-developed software \galight, which is a \texttt{python}-based open-source package that provides various astronomical data processing tools and performs 2D profile fitting. It utilizes the image modeling capabilities of \lenstronomy~\citep{2018PDU....22..189B, 2021JOSS....6.3283B} while redesigning the user interface to allow for an automated fitting ability. \galight\ uses the Particle Swarm Optimization~\citep{Kennedy1995} technique embedded in \lenstronomy\ to perform for $\chi ^2$ minimization. 

We use \galight\ to prepare our modeling ingredients, which include the following: 1) the image cutouts that cover sufficient light emission from the quasar, 2) the noise level map that describes the uncertainty of each pixel, 3) a PSF that is taken from our PSF library. In general, the noise level includes both Poisson and random background noise. Poisson noise is estimated by calculating the effective exposure time based on the \texttt{WHT} array maps and taking into account the gain values. The background RMS noise level is measured using pixels from a blank region close to the target. 

Having prepared the inputs to \galight, we perform the quasar-host decomposition. We assume the central quasar is described as a scaled PSF at an arbitrary position. We use a 2D \sersic\ profile~\citep{sersic} to model the host galaxy. If any object happens to be close to our target, we use another \sersic\ profile to model their light and remove any potential contamination from their extended profile. To avoid any unphysical results, we limit the \sersic\ parameters as follows: effective radius \reff~$\in[0\farcs{}06,2\farcs{0}]$, \sersic\ $n\in[0.3,9]$.

We recognize that the morphology of the quasar host is more complicated than a single \sersic\ profile. Nevertheless, the extended profile by a single \sersic\ model provides a good first-order approximation of the surface brightness distribution, which is sufficient to describe the host and separate it from the central quasar. The routine also produces a `data minus quasar' image, i.e., a host galaxy image, which can be used to perform a more sophisticated study of the host photometry and morphology for different scientific purposes. We apply the same \sersic\ model settings for \hst\ data and \jwst\ LW data. During the fitting, the central quasar and all galaxies within the FOV of the cutout image are modeled simultaneously since object profiles can overlap with each other.

The performance of each PSF in the library usually behaves differently for each target. Thus, we evaluate the performance of each PSF based on the final model reduced $\chi^2$ values and rank the PSFs. For \jwst, we are able to use the top 3, 5, and 8 PSFs with good performance and stack them to get a combined averaged PSF using \texttt{psfr}\footnote{\url{https://github.com/sibirrer/psfr}\label{note1}} (Birrer et al. in prep) and re-run the fitting. In the final step, we visually check the fit results for all PSFs and remove those with strong PSF mismatch residuals that could affect the result. 

Our reported measurements are based on either the top-ranked individual PSFs or the combined PSFs. We then assess measurement uncertainties by calculating the dispersion from the top-5 high-ranked PSFs. After a visual check, we find that only one PSF from the library in the F444W band can produce the clean quasar subtraction for AEGIS~482 and SDSS1419+5254; thus, the uncertainty for these two targets with this filter is not estimated. As shown later, we use 0.4 mag for the error budget for the SED fitting in the F444W bands of these two systems.

\begin{figure*}
\centering
\begin{tabular}{c c}
\hspace*{-1cm}  
{\includegraphics[trim={0cm 0cm 2.4cm 0cm},clip,height=0.65\columnwidth]{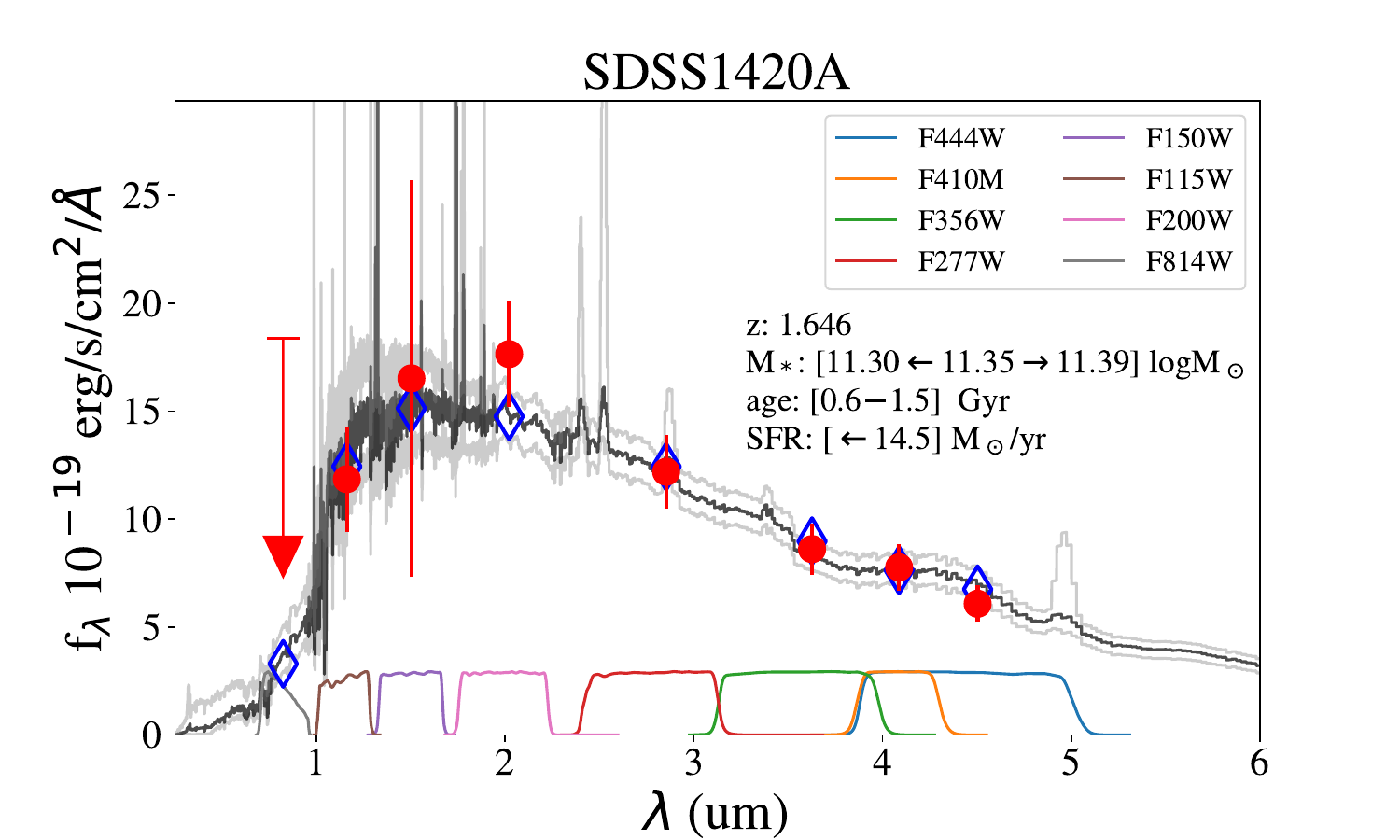}}&
{\includegraphics[trim={2.4cm 0cm 2.4cm 0cm},clip,height=0.65\columnwidth]{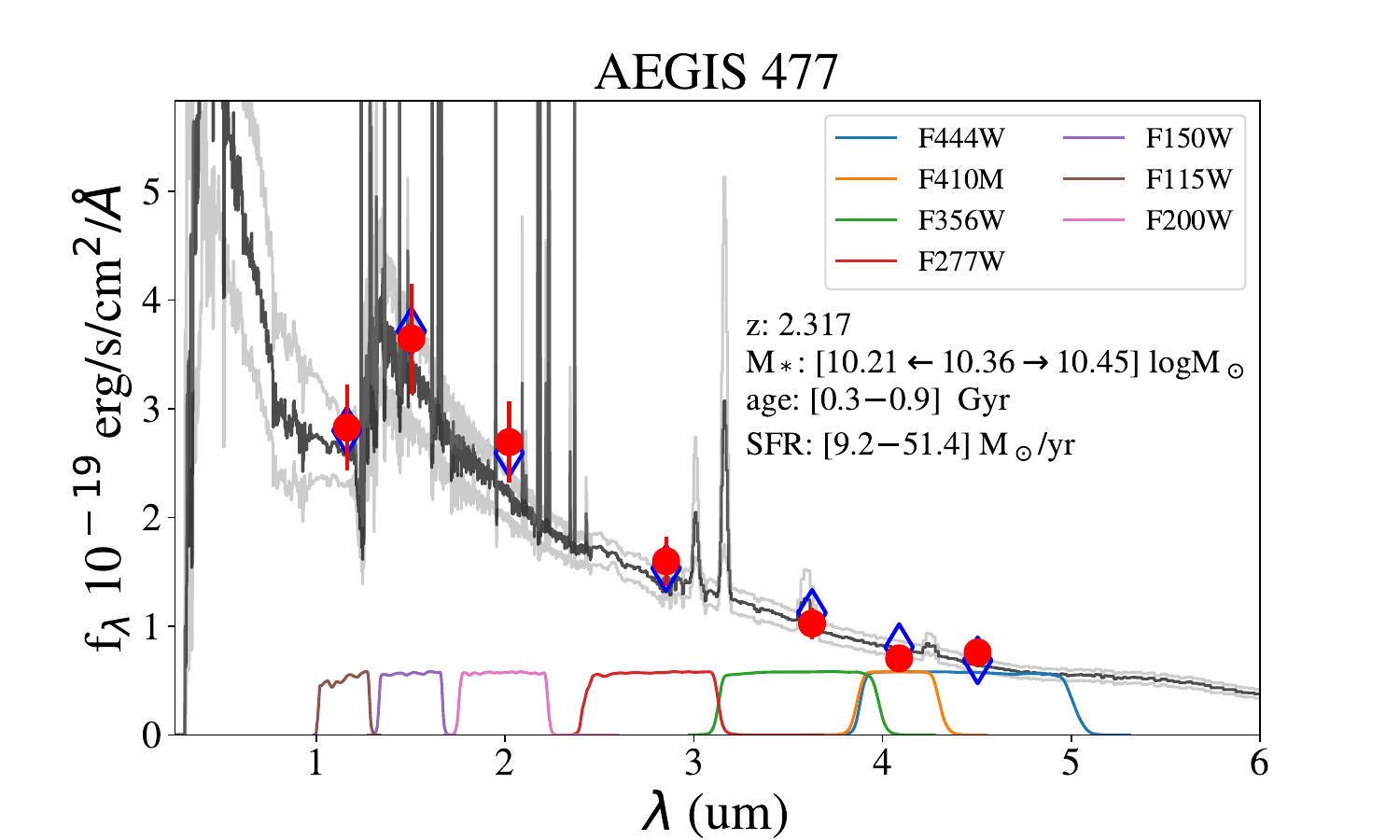}}\\
\hspace*{-1cm}  
{\includegraphics[trim={0cm 0cm 2.4cm 0cm},clip,height=0.65\columnwidth]{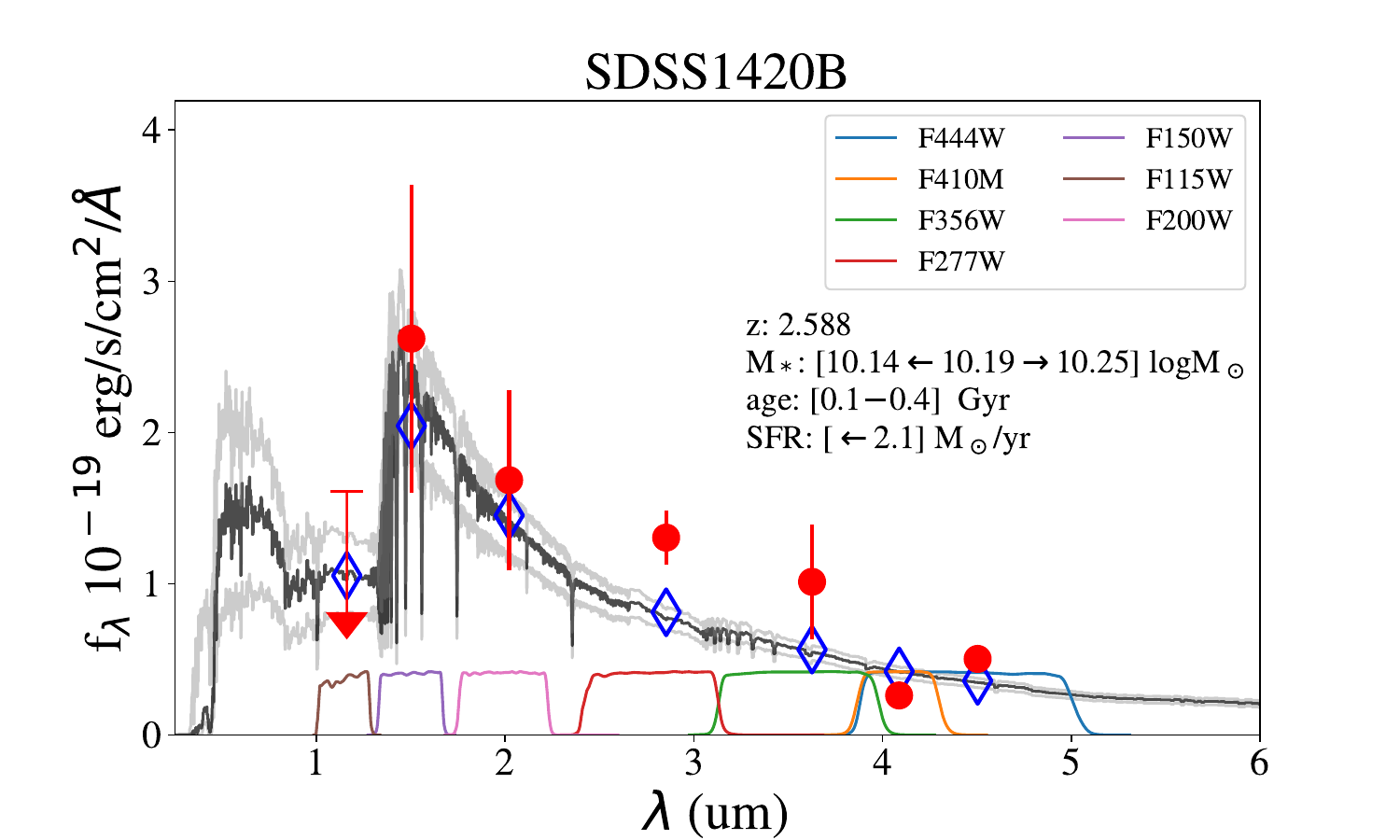}}&
{\includegraphics[trim={2.0cm 0cm 2.4cm 0cm},clip,height=0.65\columnwidth]{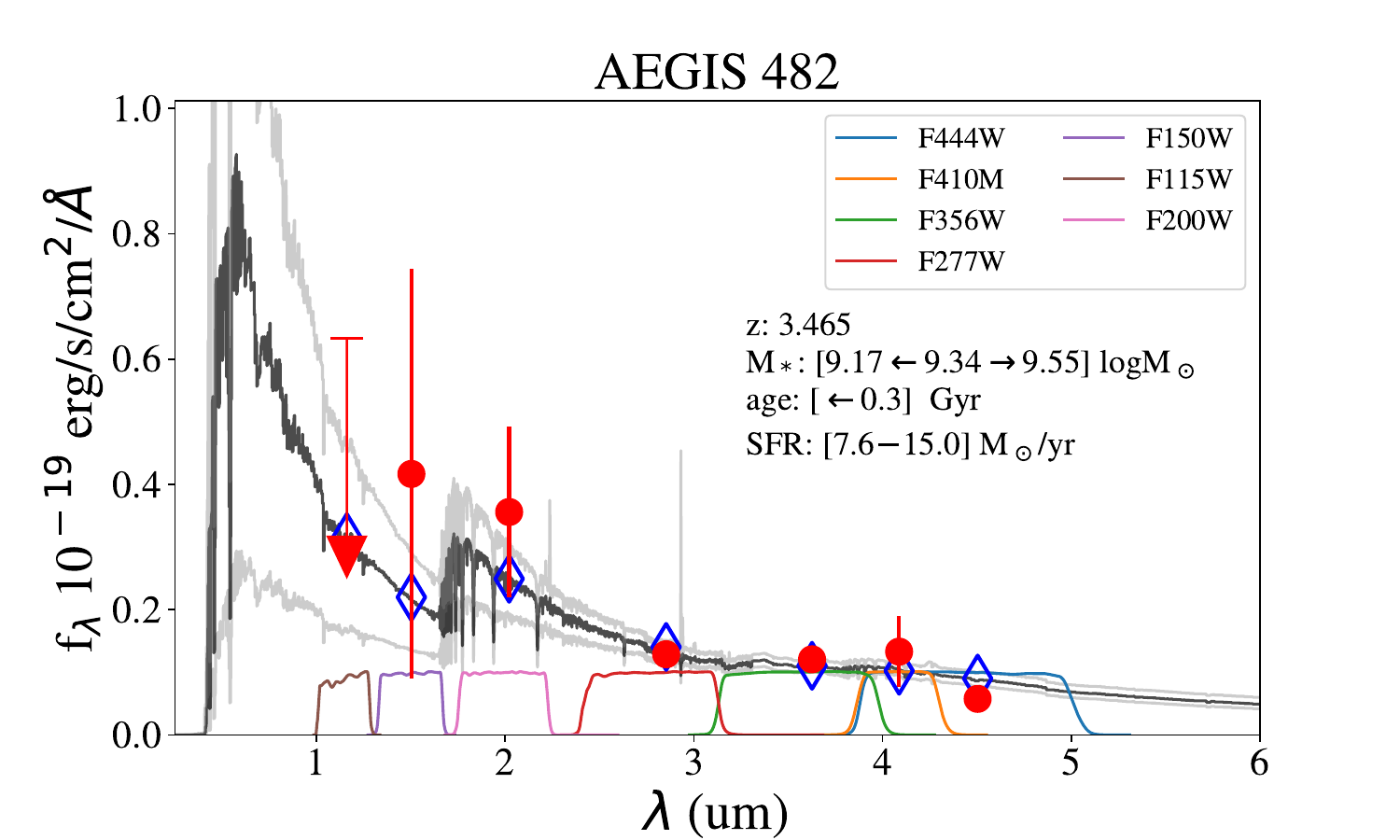}}\\
\end{tabular}
\caption{SED fitting of the host galaxy photometry based on \texttt{gsf}~\citep{Morishita2019} software. The red data points with errors indicate the host flux only, i.e., the quasar emission has been removed. The red arrow represents an upper limit. The blue diamonds show the predictions using the best-fit model. The inferred \smass\ is also shown in the figure with 16\%, 50\%, and 84\% confidence levels, along with the age and SFRs, both with lower and upper limits.
\label{fig:sed_spec}}
\end{figure*}

\section{Results} \label{sec:result}
We detect the host galaxies for all five quasars in most of the \jwst\ filters based on the 2D model fitting as described in Section~\ref{subsec:decomp}. Considering all targets and filters, the host-to-total flux ratio (total = host + quasar) spans from 5\% to 60\%. For the two quasars at $z>3$, the host ratio is lower than 10\%, which is very challenging to achieve with \hst\ and impossible at $z>3$. The stability of the \jwst\ PSFs allows a significant host detection down to these low flux ratios after the removal of the quasar component. We find that the measurements of the host properties are generally consistent across the different filters, which strengthens the fidelity of our results. We list our host measurements in Table~\ref{table:sample}.

For quasars at $z>2$, we show the images resulting from the decomposition in Figure~\ref{fig:datahost} for the F200W and F444W filters. The other filters have comparable results and are not shown due to space limitations. First, it is clearly demonstrated that any information on the host properties can only be determined after the accurate removal of the unresolved quasar component (2nd and 4th columns: host only = data $-$ quasar model component). The host galaxies of these quasars are generally isolated and do not show significant signs of undergoing a merger which is consistent with the previous studies~\citep[e.g.,][]{Cisternas2011,Kocevski2012,Ding2020} including recently reported results from CEERS on X-ray selected AGN \citep{Kocevski2022}.  We note that the fits to the \hst\ images (not shown), for this sample at $z>2$, present a larger PSF mismatch in the center that contributes significantly to the unreliability of the detection of the host.

For SDSS1420A ($z=1.646$), the 2D fitting results are presented in Figure~\ref{fig:fitting} to highlight the performance of \jwst\ (F150W, F444W) compared to \hst\ (F814W and F160W). With the 4000~\angstrom\ break ($\sim 10,000$~\angstrom\ at $z\sim1.6$) being redshifted out of the F814W filter, a very marginal detection of the host is obtained with \hst/F814W. At longer wavelengths, a highly significant detection of the host galaxy with \hst/F160W leads to very consistent fitting results with that from \jwst/F150W (see Table~\ref{table:sample}). Now, with the high-resolution power of \jwst, we can now clearly see that the quasar host is residing in a spiral galaxy with a prominent bar. Thus, the galaxy can be classified as an SBa or SBb in the Hubble classification scheme.

Overall, the majority (4/5) of quasars hosts have a low \sersic\ index ($n<2$), implying a significant disk-like component in the host galaxy. This is consistent with similar studies at lower redshift~\citep[e.g.,][]{Schawinski2012,Ding2020,Li2021,Zhuang2022}.

\subsection{SED fitting and stellar population}

We use the \texttt{gsf} package~\citep{Morishita2019} to fit the broad-band spectral energy distributions (SEDs) to infer the host stellar mass and further properties of the stellar population (i.e., stellar age and SFR). 
Stellar templates of different ages are combined to generate galaxy SED templates. In this process, weights are assigned to each stellar template and fit as free parameters to produce the final galaxy templates.
The stellar metallicity is fixed and set to the solar value. Despite the well-known degeneracy between age and metallicity, there is little effect on the determination of the \smass. The star formation rate and age will then be obtained from the final best-fit template. Dust attenuation and stellar emission lines are also considered in our SED fitting model. Note that the seven JWST filters together provide continuous wavelength coverage (see the filters' response in Figure~\ref{fig:sed_spec}) and most of the well-known rest-frame optical emission lines fall within these filters. Nevertheless, most of the filters are wide; thus, these emission lines have limited effect on our SED fitting. Indeed, we checked this by ignoring the emission lines and re-running the SED fitting which resulted in almost identical stellar properties.

We only use the host magnitudes obtained by \jwst\ filters to perform our SED fitting since the reliability of host magnitudes from \hst\ is highly uncertain. 
The SED fitting also requires errors on the magnitudes, which are estimated using the scatter of the magnitude inference from the top-5 ranked PSFs fits. If the scatter is below 0.15 mag, we fix the uncertainty at this value to account for a reported systematic error.
For AEGIS 482 and SDSS1419+5254, we adopt 0.4 mag for the error for F444W filter, since these results are provided by a single PSF.
During the SED fitting process, these errors will be used to assess the posterior probability distribution for SED parameters such as ages, stellar mass, and dust attenuation, and to perform a Markov Chain Monte Carlo routine to constrain the likelihood of these parameters.
For our sample at $z>2$, the seven \jwst\ filters span wavelengths above and below the 4000~\angstrom\ break; thus, the SED fitting should provide a reliable best-fit galaxy template. Unfortunately, SDSS1419 currently only has observations with F200W and F444W, both above the 4000~\angstrom\ break. The SED fits are presented in Figure~\ref{fig:sed_spec} with the exception of SDSS1419. We note that SDSS1420A at $z=1.646$ has a marginal detection with ACS/F814W. As a constraint, we use five times the inferred flux from the F814W filter to set up an upper limit on the SED fit to constrain the strength of the 4000~\angstrom\ break. The host stellar inference, including \smass, SFR, and age, is given at the end of Table~\ref{table:sample}. 

We classify the quasar hosts into star-forming and quiescent galaxies using the rest-frame U-V and V-J colors inferred by our model SED fits as done in \citet{vdW+2014}. For reference, we plot galaxy samples from CANDELS~\citep{vdW+2012, vdW+2014} at the equivalent redshifts to our sample (i.e., $1.6<z<3.5$), see Figure~\ref{fig:sizemass} - left. The distribution shows that the quasar hosts are mainly located in a star-forming region. One quasar host is at the boundary, possibly indicating that the host galaxy is close to transitioning to a more quiescent state, likely given its high stellar mass ($\sim2\times10^{11}$ M$_{\odot}$). Broadly, these results are consistent with numerous studies that demonstrate an association between AGN/quasar activity and ongoing star formation over a wide range of cosmic time.

\subsection{Galaxy size - stellar mass relation}

As recently reported, the stellar sizes (i.e., half-light radii) of quasars hosts span a broad distribution, with the average being intermediate between that of typical star-forming and quiescent galaxy populations. These results have been demonstrated at lower redshifts with \hst~\citep[$z\sim1.5$,][]{Silverman2019} and Subaru's Hyper Suprime-Cam Strategic Survey Program~\citep[$z < 1$,][]{Li2021}. The intermediate sizes of quasar hosts may indicate that rapidly accreting SMBHs are currently in the process of forming their central mass concentrations (i.e., bulges) in order to align with the local black hole-host mass relation. In fact, a high AGN fraction is seen in compact star-forming galaxies at $z\sim2$ \citep{Kocevski2017}.

\begin{figure*}
\centering
\begin{tabular}{c c}
\hspace*{-1cm}  
{\includegraphics[trim={0.6cm 1.1cm 2cm 2.5cm},clip, height=0.4\textwidth]{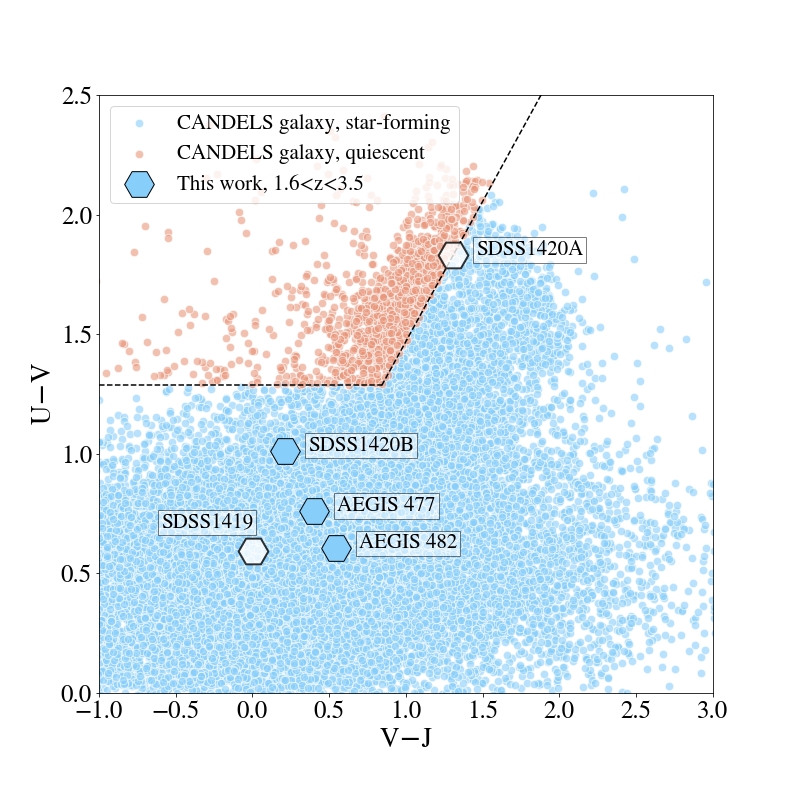}}&
{\includegraphics[trim={1cm 1.1cm 2cm 2.5cm},clip, height=0.4\textwidth]{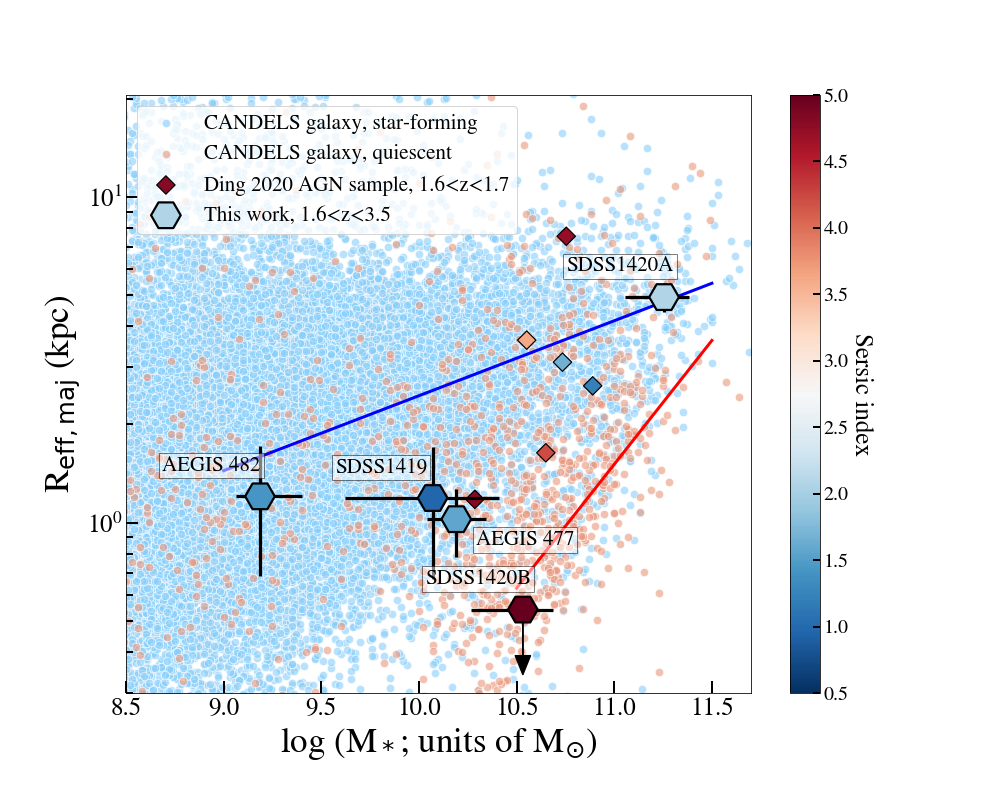}}\\
\end{tabular}
\caption{$Left:$ UVJ diagram for the five quasar systems, together with the sample from CANDELS at $1.6<z<3.5$. Two systems whose 4000~\angstrom\ break is not constrained by \jwst\ are filled in white.
$Right:$ Galaxy size--\smass\ distribution and relations. The \sersic\ index values are presented by the symbol color. For SDSS1420B, \reff\ is labeled using an arrow to indicate an upper limit. We also show the best-fit relation reported in~\citet{vdW+2014} for the star-forming (blue line) and quiescent (red line) at $z\sim2.2$. We also include six broad-line AGNs at $1.6<z<1.7$ from~\citet{Ding2020} based on \hst/WFC3 imaging. 
\label{fig:sizemass}}
\end{figure*}

Here, we investigate the location of our five high-$z$ quasar hosts relative to the size -- \smass\ relation for inactive galaxies at their respective redshift and stellar mass. The size is the median value of those measured in each filter which removes the possibility that errant fits in a single filter affect the final size measurement. As seen in Table~\ref{table:sample}, the fit using \jwst/F150W for AEGIS-482 has a size that is very small and inconsistent with the other bands. For sizes, i.e., \reff, we use recipes in~\citet{vdW+2014} to the adjust the median \reff\ to the rest-frame 5000~\angstrom. For the control sample, we use the size measurements of individual galaxies at equivalent redshifts (i.e., $1.6<z<3.5$) and the best-fit relation at $z\sim2.2$ in~\citet{vdW+2014} for star-forming and quiescent galaxies, which are classified using UVJ photometry. 

In Figure~\ref{fig:sizemass}, we show that most (4/5) quasar hosts are compact with sizes \reff$\lesssim$1~kpc. SDSS1420A is larger due to its high stellar mass. Two quasars (AEGIS~482, SDSS1420A) are in agreement with the average star-forming galaxy at their respective stellar mass, as indicated by the solid blue line. The remaining three are smaller than typical star-forming galaxies. SDSS1420B is very compact based on our upper limit resulting from a fit that hits the boundary of our allowed range in size; even so, the size constraint is consistent with quiescent galaxies. Both AEGIS~477 and SDSS1419 have host sizes ($\sim$1 kpc) right at the boundary between the two galaxy populations. Considering the quasars with \jwst\ measurements and published results using \hst/WFC3 at $1.6<z<1.7$~\citep{Ding2020}, we conclude that many of the quasar hosts are likely undergoing structural changes. The quasars with size measurements from \jwst\ are consistent with the aforementioned studies. 
Our result is consistent with the recent study using a sample lensed quasars using ALMA at $1.5<z<2.8$ observation~\citep{Stacey2021}, in which the quasar hosts are found to be more compact than normal dusty star-forming galaxies.
However, a larger sample is required to make definitive claims at $z>2$ and over a broad range in stellar mass. As evident, the \jwst\ hosts start to fill in the lower mass regime ($\log~M_*<10.5$), even down to \smass$\sim10^9$ M$_{\odot}$.    
It is worth noting that the recent work by~\citet{Miller2022} also used CEERS JWST data and identified a population of star-forming galaxies that have an SFR gradient implying centrally concentrated star-formation.

\begin{figure*}
\centering
\includegraphics[trim={4.8cm 1.5cm 21.5cm 2cm},clip,height=0.5\columnwidth]{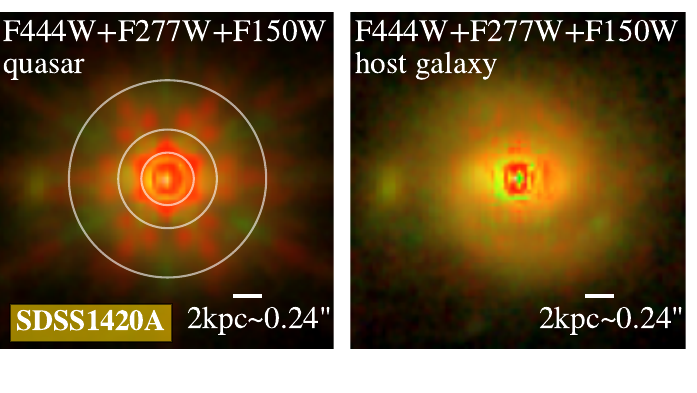}\\
\includegraphics[trim={16.6cm 0.2cm 4.2cm 1.5cm},clip,height=0.65\columnwidth]{figures/SED_map.pdf}
\caption{Spatially-resolved host properties of SDSS1420A. The top panels are ($Left$) the original quasar color image from combining three \jwst\ filters as indicated, ($Right$) the color image of the host galaxy after the central quasar is subtracted in each filter. The bottom panels from left to right are (1) the stellar mass map. (2) SFR map, and (3) age map. In various panels, three regions are indicated with a radius of 2~kpc, 4~kpc, and 8~kpc. The 2~kpc region is applied as a mask in the bottom panels.}
\label{fig:sedmap}
\end{figure*}

\subsection{SDSS1420A: spatially-resolved host properties}

SDSS1420A is an ideal laboratory to further study high-$z$ quasar hosts since spatially-resolved properties of the stellar population are feasible due to the angular extent of the host, being face-on, and the spatial resolution of multi-band \jwst\ observations. SDSS1420A is a massive disk galaxy ($log$~\smass$\sim11.3$), thus likely to be fully quenched on a short timescale and transition to a bulge-dominated galaxy. Here, our aim is to demonstrate new science capabilities enabled by \jwst. While larger samples will be required to make substantive claims, the CEERS observations of a small quasar sample can already provide further insight into the resolved properties of quasar hosts that have yet to be explored at this redshift and resolution.

As mentioned above, the decomposition of SDSS1420A reveals a clear spiral galaxy for which 2D SED fitting is feasible. We use the host-only image in seven \jwst\ filters and run the SED fitting on a pixel-by-pixel basis to generate spatially-resolved maps. Since the \hst\ F814W filter cannot resolve the host, we do not include this filter in the 2D SED fitting. To enhance the signal-to-noise and generate a map at the resolution of the PSF, we re-bin the pixels to 2$\times$2 for LW filter image and 4$\times$4 for SW filter image. Finally, the 2D SED fitting is performed at a pixel scale of $\sim0\farcs{}06$.

The final 2D SED fitting result is shown in Figure~\ref{fig:sedmap} which includes the maps of stellar mass, SFR, and galaxy age. Since residual PSF features caused by the mismatch are strong in the center pixels, we only consider the region beyond $0\farcs{}24$ ($\sim$2~kpc), which are left unmasked in the images. The 2D SED fitting result shows that the inner region has higher SFR and a younger age. 
Within an inner annulus (2.0 $<$ radius $<$ 4.0 kpc), we measure an average SFR density of 0.54~M$_{\odot}$/yr/kpc$^2$ and age of 1.26 Gyr. This SFR density is close to that of normal star-forming disk galaxies (i.e., sBzK) at $z\sim1.5$ \citep{Daddi2010}. In a larger annulus (4.0 $<$ radius $<$ 8.0 kpc), the SFR density drops to 0.095~M$_{\odot}$/yr/kpc$^2$ while the age rises to 1.64 Gyr. The relative SFR density decreases by a factor of six. Furthermore, the specific star formation rate (sSFR) is $\log$(sSFR/yr$^{-1})=-9.98$ and  $\log$(sSFR/yr$^{-1})=-10.20$ in the inner and outer annulus, respectively.
Based on our 2D SED fit results, stars are growing in the center region, possibly contributing to the build-up of the bulge component. This may argue against simple AGN quenching since one would expect less star formation in the central regions as compared to the outskirts.

\section{Concluding Remarks}\label{sec:conclusion}

We presented two-dimensional image decomposition in the optical and infrared of five luminous quasars from SDSS and AEGIS with $1.6<z<3.5$ based on \jwst+\hst\ imaging (12 filters in total from \hst/ACS/F606W to \jwst/NIRCam/F444W) provided by the CEERS collaboration. Our purpose is to detect their host galaxies and measure their properties (i.e., size, \sersic\ index, magnitude, color) for the first time beyond $z>3$. For this, we constructed a PSF library from stars within each field-of-view for every filter (see Section~\ref{subsec:decomp}). Two-dimensional modeling is then carried out using the tool \galight. A single top-ranked PSF, assessed by the goodness-of-fit, is used for host-quasar decomposition, while additional fits using other high-ranked PSFs provide an assessment of the uncertainty in the host magnitudes. With multi-band photometry of the host, we performed SED fitting to infer the properties of the stellar population, including stellar mass, age, and star formation rate.

We detect the host galaxy and model the emission for all five quasars as shown in Figure~\ref{fig:datahost} and~\ref{fig:fitting}. The host-to-total flux ratios vary greatly from 5\% to 60\% with the lower ratios only possible with the remarkable stability of the \jwst\ PSF. Our main science results are summarized as follows:

\begin{enumerate}

\item Overall, the inferred \sersic\ indices are low, with 4/5 of them having $n<2$ that agrees with past studies which demonstrate the presence of a significant stellar disk \citep[e.g.,][]{Li2021, Zhuang2022}. 

\item Based on SED fitting, the \smass\ distribution ranges from 9.2 to 11.5 in units of $\log~$M$_{\odot}$. According to the UVJ classification (Figure~\ref{fig:sizemass}, left panel), the quasar hosts lie in the star-forming region. One is close to the boundary with quiescent galaxies, possibly indicating a transitioning stage.

\item The stellar sizes cover a broad range with 2/5 consistent with the size--mass relation of star-forming galaxies, while the other three have compact sizes in agreement with quiescent galaxies or transitioning to that state. 

\item Based on a spatially-resolved investigation, SDSS1420+5300A ($z=1.646$) is a face-on galaxy with an extended profile exhibiting spiral arms and a bar. We are able to perform a 2D SED fitting using the host image with seven \jwst\ filters (Figure~\ref{fig:sedmap}). The SED map shows that the averaged SFR density in the inner annulus (2.0$<$ radius $<$ 4.0 kpc) is higher than that in an outer annulus (4.0$<$ radius $<$ 8.0 kpc) by a factor of six. The ongoing star formation and newly-formed stars in the central region are likely signs of a bulge in formation without direct evidence for quasar-mode feedback. 
\end{enumerate}

Our work demonstrates the unprecedented ability of \jwst\ to reveal the host of luminous quasars, especially at $z>3$. As demonstrated, accurate characterization of the PSF is crucial using stars within the \jwst\ FOV. The PSF shape also varies at each AGN's location; using just one empirical PSF may not always provide a decent model. Finally, the model PSF by \texttt{webbpsf} is always too peaked in the center for quasar subtraction as shown in Figure~\ref{fig:PSFprofile}.

\jwst\ has ushered in a new era for which we are now able to study quasar hosts to even higher redshift ($z\sim6$ and beyond) and lower mass. This is crucial to probe epochs closer to the formation of massive galaxies and their supermassive BHs, thus better constraining theoretical models of their connection \citep{Volonteri2021, Ding2022}.

\begin{table*}
\caption{Quasar information and inferred host properties}\label{table:sample}
\resizebox{17cm}{!}{
\begin{tabular}{llllll}
\hline
Inference & SDSS1420+5300A & AEGIS 477 & SDSS1420+5300B & SDSS1419+5254 & AEGIS 482 \\
\hline\hline
RA & 215.0233 & 214.8707 & 215.0359 & 214.9316 & 214.7552 \\
Dec & 53.0102 & 52.8331 & 53.0011 &  52.9087 & 52.8368  \\
Redshift &  1.646 & 2.317 & 2.588 & 3.442 & 3.465 \\\cline{2-6}

& \multicolumn{5}{c}{\hst\ WFC3 F160W (18 PSFs in library)} \\ \cline{2-6}
host-total flux ratio & 31.7\%$\pm$1.3\% & 20.3\%$\pm$6.4\% & 21.8\%$\pm$2.3\% & 10.7\%$\pm$1.1\% & 5.5\%$\pm$2.5\%\\
Reff ($\arcsec$) & 0.532$\pm$0.035 & 0.177$\pm$0.021 & 0.139$\pm$0.013 & 0.185$\pm$0.009 & 0.142$\pm$0.030\\
Reff (kpc) & 4.50$\pm$0.30 & 1.45$\pm$0.17 & 1.11$\pm$0.11 & 1.36$\pm$0.07 & 1.04$\pm$0.22\\
\sersic\ $n$ & 2.2$\pm$0.7 & 0.4$\pm$0.1 & 0.3$\downarrow$ & 0.3$\downarrow$ & 0.3$\downarrow$\\
host mag & 20.96$\pm$0.15 & 23.49$\pm$0.26 & 22.87$\pm$0.15 & 23.42$\pm$0.15 & 25.11$\pm$0.34\\
total mag & 19.715$\pm$0.001 & 21.758$\pm$0.005 & 21.216$\pm$0.008 & 20.995$\pm$0.007 & 21.958$\pm$0.006\\
$\chi ^2$ (Reduced) & 15.04 & 2.41 & 4.39 & 5.59 & 2.68\\
\cline{2-6}

& \multicolumn{5}{c}{\jwst\ NIRCam F150W (26 PSFs in library)} \\ \cline{2-6}
host-total flux ratio & 32.1\%$\pm$8.8\% & 39.8\%$\pm$6.1\% & 14.1\%$\pm$3.9\% & \nodata & 5.8\%$\pm$2.2\%\\
Reff ($\arcsec$) & 0.478$\pm$0.07 & 0.101$\pm$0.016 & 0.06$\downarrow$ & \nodata & 0.091$\pm$0.031\\
Reff (kpc) & 4.05$\pm$0.59 & 0.83$\pm$0.13 & 0.48$\downarrow$ & \nodata & 0.67$\pm$0.23\\
\sersic\ $n$ & 2.0$\pm$0.8 & 1.7$\pm$1.2 & 9.0$\uparrow$ & \nodata & 7.1$\pm$1.9\\
host mag & 21.17$\pm$0.58 & 22.81$\pm$0.15 & 23.17$\pm$0.41 & \nodata & 25.16$\pm$0.78\\
total mag & 19.935$\pm$0.142 & 21.808$\pm$0.008 & 21.042$\pm$0.003 & \nodata & 22.068$\pm$0.003\\
$\chi ^2$ (Reduced) & 3.24 & 3.34 & 2.22 & \nodata & 1.29\\
\cline{2-6}
& \multicolumn{5}{c}{\jwst\ NIRCam F200W (23 PSFs in library)} \\ \cline{2-6}
host-total flux ratio & 41.4\%$\pm$0.3\% & 36.5\%$\pm$1.1\% & 15.0\%$\pm$3.7\% & 16.3\%$\pm$5.5\% & 8.3\%$\pm$4.7\%\\
Reff ($\arcsec$) & 0.436$\pm$0.004 & 0.122$\pm$0.003 & 0.089$\pm$0.029 & 0.161$\pm$0.03 & 0.159$\pm$0.045\\
Reff (kpc) & 3.69$\pm$0.04 & 1.00$\pm$0.03 & 0.71$\pm$0.23 & 1.18$\pm$0.22 & 1.17$\pm$0.33\\
\sersic\ $n$ & 2.3$\pm$0.1 & 1.2$\pm$0.1 & 7.7$\pm$1.3 & 1.2$\pm$1.9 & 1.4$\pm$3.6\\
host mag & 20.45$\pm$0.15 & 22.49$\pm$0.15 & 23.0$\pm$0.38 & 23.01$\pm$0.26 & 24.69$\pm$0.41\\
total mag & 19.497$\pm$0.002 & 21.401$\pm$0.003 & 20.947$\pm$0.005 & 21.043$\pm$0.002 & 21.995$\pm$0.004\\
$\chi ^2$ (Reduced) & 2.74 & 2.69 & 2.17 & 5.42 & 1.58\\
\cline{2-6}

& \multicolumn{5}{c}{\jwst\ NIRCam F356W (12 PSFs in library)} \\ \cline{2-6}
host-total flux ratio & 40.0\%$\pm$0.2\% & 44.4\%$\pm$2.2\% & 29.3\%$\pm$6.8\% & \nodata & 10.9\%$\pm$0.6\%\\
Reff ($\arcsec$) & 0.4000$\pm$0.004 & 0.129$\pm$0.008 & 0.095$\pm$0.035 & \nodata & 0.221$\pm$0.007\\
Reff (kpc) & 3.39$\pm$0.04 & 1.06$\pm$0.06 & 0.76$\pm$0.28 & \nodata & 1.63$\pm$0.05\\
\sersic\ $n$ & 2.1$\pm$0.1 & 1.2$\pm$0.1 & 5.6$\pm$2.3 & \nodata & 1.3$\pm$0.2\\
host mag & 19.96$\pm$0.15 & 22.27$\pm$0.15 & 22.29$\pm$0.40 & \nodata & 24.6$\pm$0.15\\
total mag & 18.969$\pm$0.006 & 21.391$\pm$0.001 & 20.953$\pm$0.004 & \nodata & 22.186$\pm$0.001\\
$\chi ^2$ (Reduced) & 17.36 & 1.11 & 3.56 & \nodata & 3.53\\
\cline{2-6}

& \multicolumn{5}{c}{\jwst\ NIRCam F444W (18 PSFs in library)} \\ \cline{2-6}
host-total flux ratio & 32.8\%$\pm$1.1\% & 48.1\%$\pm$4.9\% & 22.6\%$\pm$2.4\% & 12.1\%& 6.2\%\\
Reff ($\arcsec$) & 0.386$\pm$0.008 & 0.102$\pm$0.011 & 0.067$\pm$0.013 & 0.312 & 0.295\\
Reff (kpc) & 3.27$\pm$0.07 & 0.84$\pm$0.09 & 0.54$\pm$0.10 & 2.30 & 2.17\\
\sersic\ $n$ & 1.9$\pm$0.1 & 1.6$\pm$0.2 & 4.9$\pm$0.6 & 0.7 & 0.4\\
host mag & 19.87$\pm$0.15 & 22.13$\pm$0.17 & 22.58$\pm$0.15 & 23.44 & 24.93\\
total mag & 18.548$\pm$0.032 & 21.245$\pm$0.039 & 20.83$\pm$0.016 & 21.005$\pm$0.010 & 21.648$\pm$0.009\\
$\chi ^2$ (Reduced) & 10.45 & 3.27 & 2.03 & 12.13 & 4.21\\
\cline{2-6}

& \multicolumn{5}{c}{SED inference} \\ \cline{2-6}
$M_*$ ($\log$ M$_{\odot}$) & [11.30, 11.35 ,11.39] & [10.21, 10.36 ,10.45] & [10.14, 10.19 ,10.25] & [9.02, 10.16 ,10.41] & [9.17, 9.34 ,9.55] \\
age (Gyr)  &                  [0.6$-$1.5]  & [0.3$-$0.9] & [0.1$-$0.4] & [0.1$-$0.5] & [$\leftarrow$0.3] \\
SFR (M$_{\odot}$/yr) &       [$\leftarrow$14.5] & [9.2$-$51.4]  & [$\leftarrow$2.1] & [$\leftarrow$6.4] & [7.6$-$15.0] \\
\cline{2-6} \\
\hline
\hline
\end{tabular}}
\tablecomments{The information of each target is listed by column. For page limitation, we only present one band from \hst\ and four bands from \jwst. The total number of selected PSF from the entire FOV for each filter is also shown in the table. The \reff\ is referred to as the semi-major-axis half-light radius. We present the inferred values for the SED fitting inference at 16\%, 50\% (if present), and 84\% confidence. For AEGIS~482 and SDSS1419+5254, only one PSF in the library can produce a decent fit in the F444W filter (see Figure~\ref{fig:datahost}); we thus only show the result using that PSF. The up/down arrows are used to indicate the \sersic\ fitting hit the parameter boundary. Note that, for $z>2$ sample, the host measurements provided by \hst\ should be considered highly uncertain.
The inference of the total (i.e., host+quasar) magnitudes is also provided. Combining with host magnitude, the quasar properties (e.g., magnitude, luminosity) can be obtained.
}
\end{table*}

\section*{Acknowledgments}
\newpage 
We thank the CEERS collaboration that achieves the \jwst\ observation for the data and for developing their observing program with a zero-exclusive-access period. We thank Takahiro Morishita for the help with using the \texttt{gsf} package. We thank Simon Birrer for providing the \texttt{psfr} package that has been used for the PSF stacking. We thank Connor Bottrell for providing the algorithm for momentum measurement that improves the \galight\ fitting performance. We thank Knud Jahnke, Takuma Izumi, Yoshiki Matsuoka, Lilan Yang, and Hassen Yesuf for the useful discussion.

Quasar decomposition works were performed on the {\it idark} cluster at Kavli IPMU, The University of Tokyo.

All of the data presented in this paper were obtained from the Mikulski Archive for Space Telescopes (MAST) at the Space Telescope Science Institute. The specific observations analyzed can be accessed via \dataset[https://doi.org/10.17909/3pf0-8b20]{https://doi.org/10.17909/3pf0-8b20}. 
STScI is operated by the Association of Universities for Research in Astronomy, Inc., under NASA contract NAS5-03127 and NAS5–26555 for JWST. Support to MAST for these data is provided by the NASA Office of Space Science via grant NAG5–7584 and by other grants and contracts.
These observations are associated with program 1345.

This work was supported by World Premier International Research Center Initiative (WPI), MEXT, Japan. X.D. is supported by JSPS KAKENHI Grant Number JP22K14071. J.S. is supported by JSPS KAKENHI (grant Nos. JP18H01251 and JP22H01262) and the World Premier International Research Center Initiative (WPI), MEXT, Japan. 
M.O. is supported by the National Natural Science Foundation of China (12150410307).

\facilities{\hst, \jwst}

\software{\galight\citep{Ding2020,Ding2021a}, \lenstronomy~\citep{2018PDU....22..189B, 2021JOSS....6.3283B}, \texttt{gsf}~\citep{Morishita2019}, \texttt{psfr}\footnoteref{note1} (Birrer et al. in prep), \texttt{astropy}~\citep{2013A&A...558A..33A,2018AJ....156..123A}, \texttt{photutils}~\citep{larry_bradley_2020_4044744}}.

\bibliographystyle{aasjournal}

\end{document}